# The solution of the Laplace equation with the Robin boundary conditions: Applications to inverse problems.

Stéphane Mottin
18 rue B. Lauras, UMR5516 CNRS, University J.Monnet, University of Lyon,
F-42000 St-Etienne, France
mottin@univ-st-etienne.fr

**Abstract**
This paper studies the following problem with azimuthal symmetry:
$\Delta u=0$ in a unit sphere
$\partial u(r, \zeta)/\partial r|_{r=1} + h\, u(1, \zeta) = f(\zeta)$ on a unit sphere,
$\Delta$ is the Laplace operator. $\zeta=\cos(\theta)$, $\theta$ is the azimuthal angle and $h \in \mathbb{R}^{*+}$.
The function $f(\zeta)$ is a prescribed function and is assumed to be a square-integrable function.
Many solutions of the boundary value problems in spherical coordinates are available in the form of infinite series of Legendre polynomials but the evaluation of the summing series shows many computational difficulties. Here, the closed-form solution of the Laplace equation with this Robin boundary conditions on a sphere is solved by the Legendre transform. In many experimental approaches, this weight « h », the Robin coefficient, is the main unknown parameter for example in transport phenomena where the Robin coefficient is the dimensionless Biot number. The usefulness of this solution is illustrated by some examples of inverse problems.

## 1. Introduction

The study of many transport mechanisms in confined biological or physical domains is often linked to elliptic partial differential equation with sphere media. The Laplace equation is arguably the most important differential equation in all of applied mathematics. It arises in an astonishing variety of physical and mathematical systems, ranging through electromagnetism, fluid mechanics, potential theory, solid mechanics, heat conduction, geometry and on and on. Laplace equation is the simplest elliptic partial differential equation modelling a plethora of steady state phenomena. The Laplacian with the Robin boundary conditions on a sphere is one of the most important boundary value problem in many sciences because spherical geometry is everywhere from the biggest structures in the universe to the smallest particles. The Robin boundary conditions imply a constant "h" and corresponds to the Dirichlet conditions (h→+∞), or to the Neumann conditions (h→0). Another way of viewing the Robin boundary conditions is that it typies physical situations where the boundary "absorbs" some, but not all, of the energy, heat, mass…, being transmitted through it.

Despite these strong interests, very few analytical solutions of the Laplace equation for a sphere are known [1, 2]: —first, the solution of the first boundary value problem is the well-known Poisson's integral for the sphere [3],  —second, the exact solution of the Neumann boundary conditions was published sixty years ago [4]. To the best of our knowledge, the analytical solution for the Robin boundary problem on a sphere is not known. Here the Laplace equation with an homogeneous isotropic medium and the axisymmetric sphere problem are considered with the most general nonhomogeneous Robin boundary conditions.

In mathematics the generalized Robin problem (where h is a continuous function) for the Laplace equation is still a work in progress [5, 6]. Research of method for solving second-order elliptic differential equations subject to the nonhomogeneous Robin boundary conditions is also under progress [7-9].

The Laplace equation is a special case of the Helmholtz equation [10]:

$$\Delta u(\mathbf{r}) + K(\mathbf{r}) u(\mathbf{r}) = 0 \qquad (1)$$

with $\mathbf{r} \in \mathbb{R}^3$.

In physical mathematics, four cases are distinguished:
1) K(r) is all times positive: the Helmholtz equation; $K(\mathbf{r}) = k^2(\mathbf{r})$,
2) K(r) is all times negative: the diffusion-reaction equation; $K(\mathbf{r}) = -k^2(\mathbf{r})$,
3) K(r) is equal to $\gamma k^2$ ($\gamma \in \mathbb{C}$ and $|\gamma|=1$) : the "Generalized Helmholtz" Equation,
4) K(r) is equal to zero: the Laplace equation.

In the case n°2, the quantity "k" could also be analyzed as an absorption coefficient [10-12] or as a refractive index [2]… It is well-known that this class of elliptic equation is related to spherical harmonics in (p+2)-dimensional Euclidean space with p=1,2,… [13] and for the first case (p=1), the Gegenbauer polynomials are the Legendre polynomials.

If the medium is considered to be not homogeneous (propagation in a dispersive medium or in a complex absorbing medium) then k(**r**) is not a constant. Li *et al* [2] published the three-dimensional analytical solution in a semi-infinite linearly inhomogeneous medium with $k=k_0 (1+\beta z)^{1/2}$ where $k_0$ and ß are positive constants and (1+ßz) is always positive. Some contributions were published with special K(r) [13]. There are a lot of works on layered media [14]. The asymptotes of the time-dependent solutions on semi-infinite medium were analyzed [14]. Recently the theory of homogenization have been applied on this problem [10, 15].

If the medium is homogeneous, k(**r**) is a constant (named k and k≠0), then the solutions are expressed as infinite series [1, 11, 16]. Several problems for elliptic equation in three spatial dimensions with Dirichlet or Neumann conditions have been solved in the interior of a sphere and of a spherical sector by the Fokas method in term of the integral representation of the solution [7].

The Laplace equation corresponds to the lossless diffusion equation and more generally when k=0 (or k→0). The solutions of Laplace equation are called harmonic functions. In this article, the method of integral transforms on finite intervals with the Legendre transform [17] will be used. This equation is the simplest representative of second-order partial differential equations of elliptic type. Laplace equation arises in the study of a plethora of physical phenomena, including electrostatic or gravitational potential, the displacement field of a two- or three-dimensional elastic membrane and the velocity potential of an incompressible fluid flow. The physical meaning of the Laplace equation is that it is satisfied by the potential of any such field in source-free domains 𝔇 of the Euclidean space $\mathbb{R}^n$ (n≥2). For example, the Laplace equation is satisfied by the potential of an electrostatic field in a domain free from charges, the gravitational potential of the gravity force in domains free from attracting masses... Thus, the Laplace equation expresses the conservation law for a potential field. It also arises in many other problems in mathematical physics in which stationary fields are considered, for example in the study of a stationary temperature distribution. The physical quantity of interest is also the field, for example the electric field $\mathbf{E} = \nabla u$ or the Temperature field… Then with an analytical solution of the potential "u", the field can be directly calculated.

The Laplace equation can be solved by separation of variables in eleven coordinate systems that the Helmholtz equation (equation (1)) can. In addition to these eleven coordinate systems, separation can be achieved in two additional coordinate systems (bispherical and toroidal coordinates), bringing the total number of separable systems for Laplace equation to thirteen [18]. Among these thirteen coordinate systems, the spherical coordinates are special because Green's function for the sphere can be used as the simplest majorant for Green's function for an arbitrary bounded domain [19].

The use of symbols differs between sources. In one system frequently encountered in physics (r, θ, φ) gives the radial distance, polar angle, and azimuthal angle, whereas in another system used in mathematics (r, θ, φ) gives the radial distance, azimuthal angle, and polar angle. Due to these misuses of these notations, the right-handed spherical coordinate system (r, θ, φ) to denote radial distance, inclination, and azimuth, respectively, as specified by ISO standard 80000-2 :2009, will be used:

θ is the angle defined by the zenith axis (z axis); 0≤ θ ≤π. It is the polar angle measured down from the "north pole".

φ is the angle defined in the plane perpendicular to the zenith axis (the xy-plane); 0≤φ<2π. We consider situations with complete rotational symmetry about the z-axis (azimuthal symmetry or axial symmetry) in order to focus on the Legendre transform.

$$\Delta u = \frac{\partial}{\partial r}\left(r^2 \frac{\partial u}{\partial r}\right) + \frac{1}{\sin\theta} \frac{\partial}{\partial \theta}\left(\sin\theta \frac{\partial u}{\partial \theta}\right) = 0 \quad (2)$$

with r < 1.

A potential independent of φ is a potential azimuthally invariant. There are many interesting systems which are more or less of this type.

The first step in solving partial differential equations using separable variables is to assume a solution of the form : u(r,θ)= R(r) G(θ) where R(r) is a function only of r, and G(θ) is a function of θ. The Laplace'equation becomes (prime denotes derivation):

$$\frac{1}{R(r)} \frac{d(r^2 R\prime(r))}{dr} + \frac{1}{G(\theta)\sin\theta} \frac{d(\sin\theta\, G\prime(\theta))}{d\theta} = 0 \quad (3)$$

Notice that the derivates in this equation (3) are no longer partial derivatives. This is because this well-known method of separable variables has produced two terms; one is solely a function of r and the other a function of θ. Equation (3) allows us to separate Laplace's equation into two separate ordinary differential equations. Each term on the right hand side of equation (3) is equal to a constant. This means we can separate equation (3) into:

$$r^2 \frac{d^2 R}{dr} + 2r \frac{dR}{dr} - \big(n(n+1)\big)R = 0 \quad (4a)$$

$$\frac{d}{d\zeta}\left((1-\zeta^2)\frac{dG}{d\zeta}\right) + \big(n(n+1)\big) G = 0 \quad (4b)$$

where the variable ζ=cos θ (-1≤ ζ ≤1) is introduced.

The separation constant is chosen n(n+1) because by writing the separation constant in this way we will produce a well known differential equation whose general solution we already know. Equation (4a) is a classic example of Cauchy-Euler equation or a fairly simple example of a Frobenius equation. Equation (4b) is the Legendre's differential equation [20]. The solutions to the Legendre equation are the Legendre polynomials by definition. Using either methods of Euler's equations or the method of Frobenius, the solution to equation (4a) is well-known: R(r)=

$A_n r^n + B_n r^{-(n+1)}$ where $A_n$ and $B_n$ are constants which will be determined once we apply specific boundary equations. The most general form that a solution can have is

$$u(r, \zeta) = \sum_{n=0}^{\infty} (A_n r^n + B_n r^{-(n+1)}) P_n(\zeta) \qquad (5)$$

The Legendre polynomials $P_n(\zeta)$ form a complete set on the interval $0 \leq \cos\theta \leq \pi$ or $-1 \leq \zeta \leq 1$. Thus any specified φ-independent potential on a spherical surface can be expressed as a sum of $P_n$'s.

## 2. The Robin boundary conditions

In the theory of linear partial differential equations, a well-posed problem consists of a differential equation subject to certain boundary conditions such that the solution is unique. The third boundary value problem is a well-posed problem [3].

Let $\Omega$ be the unit sphere domain in $\mathbb{R}^3$, $\partial\Omega$ be its surface (r=1).

Assuming azimuthal symmetry, we write:

$$\partial u(r, \zeta)/\partial r |_{r=1} + h\, u(1, \zeta) = f(\zeta) \qquad (6)$$

$h \in \mathbb{R}^{*+}$. The function $f(\zeta)$ is a prescribed function of $\zeta$ and is assumed to be a square-integrable function.

This boundary operator appear in many contexts in science and engineering, for exemple in many transport phenomena [21] or in optics [10]. Sometimes "h" is called the Biot number [22]. Robin boundary conditions are also called impedance boundary conditions in some engineering problems. They are commonly used in solving Sturm–Liouville problems which appear in many contexts in science. The third boundary condition or the Robin boundary condition is also known as Newton boundary condition [23]. A method for treating general boundary conditions in the finite element method [24] considers these general Robin boundary conditions: $\partial u/\partial n = 1/\varepsilon\,(u_0-u)-g$ with u and g two functions, and $\varepsilon \in \mathbb{R}^+$. If $\varepsilon \to 0$ then $u=u_0$ on the boundary. If $\varepsilon \to \infty$ then $\partial u \partial n = g$ on the boundary.

The problem is to find the solution of the equation (3) continuous on the closed domain ($\Omega$-$\partial\Omega$) and satisfying equation (6) on $\partial\Omega$. The solution to the Laplace equation with Dirichlet's conditions is the well-known Poisson's integral for the sphere [3, 17] and the solution of the Neumann problem for the sphere was published in the context of the use of electric images [4]. The Green's function for the Dirichlet problem and the Neumann problem for the sphere were obtained by the method of images and the inversion with respect to the sphere, which is a Kelvin Transformation [4]. The third problem is more complex than these two problems. It is not possible to extend the method of images to the third problem of the Laplace equation, and also to the steady-state Helmholtz equation [25].

## 3. The Legendre transform

The Legendre transform (more exactly the Legendre polynomial transform) and the inverse Legendre transform are respectively defined by :

$$\underline{u}(n) = Le(u(\zeta)) = \int_{-1}^{1} u(\zeta)\, P_n(\zeta)\, d\zeta \qquad (7)$$

n=0, 1, 2, 3…

Le ( ) is a linear integral transformation.

$$u(\zeta) = Le^{-1}(\underline{u}(n))(\zeta) = \sum_{n=0}^{\infty} (n+1/2)\, \underline{u}(n)\, P_n(\zeta) \qquad (8)$$

with $-1 < \zeta < 1$.

The central property of this transformation is obtained applying successive integration by parts according to the equation (4b) [20]:

$$\text{Le}\left(\frac{d}{dx}(1-x^2)\frac{dg(x)}{dx}\right) = -n(n+1)\,\underline{g}_{(n)} \tag{9}$$

This transformation replaces a differential operation by the algebraic operation $-n(n+1)\,\underline{g}_{(n)}$.
The Legendre polynomials $P_0(\zeta)$, $P_1(\zeta)$,...$P_n(\zeta)$ are solutions of the Legendre equation. More precisely, the solution of Legendre equation can be stated as:

$w(\zeta) = C_n\, P_n(\zeta) + D_n\, Q_n(\zeta)$

where $C_n$ and $D_n$ are arbitrary constants and $Q_n(\zeta)$ are the Legendre functions of the second kind. In our physical applications no singularities are present along the polar axis, we disregard the Legendre functions of the second kind which are singular for $\zeta = \pm 1$. Then $D_n$ is equal to 0.

Integrating the equation (6) with respect to $\zeta$ from $-1$ to $1$ leads to:

$$\partial \underline{u}_{(r=1,n)}/\partial r + h\,\underline{u}_{(r=1,n)} = \underline{f}_{(n)} \tag{10}$$

In the Legendre's space, equation (4) is the following:

$$\underline{u}_{(r,n)} = A_n\, r^n + B_n\, r^{-n-1} \qquad \text{with } 0 \leq r \leq 1 \tag{11}$$

Evidently, if u is finite as $r \to 0$ then $\underline{u}$ must be finite at $r=0$, and then $B_n=0$ and $A_n = \underline{f}_{(n)}/(n+h)$.

$$\underline{u}_{(r,n)} = \underline{f}_{(n)}\, r^n/(n+h) \tag{12}$$

The inverse transform leads to the following infinite serie:

$$u_{(r,\zeta)} = \text{Le}^{-1}(\underline{u}_{(r,n)})(\zeta) = \sum_{n=0}^{\infty} (n+1/2)\,\underline{f}_{(n)}\, r^n/(n+h)\, P_n(\zeta) \tag{13}$$

Many solutions of boundary value problems are available in the form of infinite series and can be computed by expanding in eigenfunctions [1]. The evaluation of the summing series of the form $\sum_{n=0}^{\infty} q_n\, P_n(\cos\theta)$ presents many computational problems [11]: $q_n$ are often slowly decaying, and $P_n$ is more and more oscillatory. This has the effect that we are subtracting two numbers of nearly equal magnitude with the attendant loss of precision. The theoretical problem of approximation such infinite series (Gibbs phenomenon) is multifaceted [26]. Legendre transform and spherical harmonic transform are the most important orthogonal function transforms only except Fourier transform, and research of fast algorithms are still under progress [27].

Now we present some properties of Legendre polynomials in regards to the generating function $(1-2\zeta r+r^2)^{-1/2}$. The most important properties is that Legendre polynomial $P_n(\zeta)$ is also defined as the coefficient of $\zeta^n$ in the expansion of $(1-2\zeta r+r^2)^{-1/2}$ in ascending powers of r [28]:

$$(1-2\zeta r+r^2)^{-1/2} = \sum_{n=0}^{\infty} (n+1/2)\left(r^n/(n+1/2)\right) P_n(\zeta) \tag{14}$$

where $r < \min |\zeta \pm i\sqrt{(1-\zeta^2)}|$ wich is always true.

And obviously, the inverse Legendre transform of $r^n/(n+1/2)$ is $(1-2\zeta r+r^2)^{-1/2}$. $R=(1-2\zeta r+r^2)^{-1/2}$ is the inverse of the distance between the two following points $(r,\theta)$ and $(1,0)$. From a mathematics point of view, this Legendre polynomial generating function provides a convenient way of deriving many useful properties. At $r=1$ and $\zeta=1$ ( $(1,0)$ in polar coordinates, the generating function R presents a singularity, but evidently its integral

$\left(\int_0^r (1-2\zeta r+r^2)^{-1/2} dr\right)$ converges for $r=\zeta=1$.

More generally, the Legendre generating function is a special case of this generating function:

$$(1-2\zeta r+r^2)^{-\beta} = \sum_{n=0}^{\infty} r^n\, C_n^{(\beta)}(\zeta) \tag{15}$$

The coefficient $C_n^{(\beta)}(\zeta)$ are the ultraspherical polynomials (proportional to the Gegenbauer polynomials) [28] and for $\beta=1/2$, this equation reduces to :

$$C_n^{(\beta)}(\zeta) = P_n(\zeta) \tag{16}$$

One of the best representation is the Gauss functions because the Legendre functions belong to Gauss hypergeometric series:

$$P_n(\zeta) = {}_2F_1(-n, 1+n; 1; 1/2-\zeta/2) \tag{17}$$

The equation (13) must be analyzed as a product of two Legendre transforms in order to obtain an inversion formula of the product of two transforms. A convolution property of the transformation is one that expresses the inverse transform of the product of two transforms in terms of the two functions without direct recourse to the basic inversion formula. More than fifty years, a theorem about the inverse transforms of products of Legendre transforms was published [3]:

If $\underline{w}(n) = Le(w(\zeta))$ and $\underline{v}(n) = Le(v(\zeta))$, then

$$\underline{w}(n)\cdot\underline{v}(n) = \int_0^\pi P_n(\lambda)\, w(\lambda)\, \sin\lambda\, d\lambda \ . \ \int_0^\pi P_n(\eta)\, v(\eta)\, \sin\eta\, d\eta$$

The product $\underline{w}(n)\cdot\underline{v}(n)$ is the transform of the function $h(x)$ which corresponds to the following convolution $w(x) * v(x)$:

$$h(\zeta) = (\pi)^{-1} \int_0^\pi W(\cos\sigma)\, \sin\sigma \int_0^\pi V(\cos\gamma)\, d\lambda\, d\sigma \tag{18}$$

with $\quad \cos\gamma = \cos\theta\,\cos\sigma + \sin\theta\,\sin\sigma\,\cos\lambda. \tag{19}$

## 4. The main result. The inverse Legendre transform of $r^n/(n+h)$

Now we will find the closed form of $\sum_{n=0}^{\infty} (n+1/2)\, r^n/(n+h)\, P_n(\zeta)$.

It could be expressed as the following sum:

$$v(r,\zeta) = \sum_{n=0}^{\infty} (n+1/2)(r^n/(n+h)) P_n(\zeta)$$
$$= v_1(r,\zeta) + 1/2\, v_2(r,\zeta) \tag{20a}$$

with these two functions:

$$v_1(r,\zeta) = \sum_{n=0}^{\infty} n\, r^n/(n+h)\, P_n(\zeta) \tag{20b}$$

$$v_2(r,\zeta) = \sum_{n=0}^{\infty} r^n/(n+h)\, P_n(\zeta) \tag{20c}$$

$v_2(r,\zeta)$ can be found by multiplying equation (14) by $r^{h-1}$ and integrating with respect to r:

$$\sum_{n=0}^{\infty} \left(\int_0^r \rho^{n+h-1}\, d\rho\right) P_n(\zeta) = \int_0^r \rho^{h-1}\, d\rho / (1-2\zeta\rho+\rho^2)^{1/2} \tag{21}$$

and $v_2(r,\zeta) = \sum_{n=0}^{\infty} r^n/(n+h)\, P_n(\zeta) = r^{-h} \int_0^r \rho^{h-1}\, (1-2\zeta\rho+\rho^2)^{-1/2}\, d\rho \tag{22}$

$\int_0^r (\rho^{h-1}\,(1-2\zeta\rho+\rho^2)^{-1/2})\, d\rho$ ) is also the inverse function of $\underline{z}(n) = r^{n+h}/((n+h)(n+1/2))$.

The definite integral (equation 22) corresponds to the main difficulty.

$$A_h(r, \cos\theta) = \int_0^r \frac{\rho^{h-1}}{(1-2r\cos\theta+r^2)^{1/2}} d\rho \qquad (23)$$

The generating function $(1-2r\cos\theta + r^2)^{1/2}$ is the distance between the points $(r,\theta)$ and $(1,0)$, then it is convenient to use the complex coordinates $e^{i\theta}$.

Let us write this definite integral:

$$B(h,r,\theta) = \int_0^1 \frac{t^{h-1}}{(1-tr\,2\cos\theta+t^2r^2)^{1/2}} dt \qquad (24)$$

$$B(h,r,\theta) = \int_0^1 \frac{t^{h-1}}{(1-tr(e^{i\theta}+e^{-i\theta})+t^2r^2)^{\frac{1}{2}}} dt$$

$$= \int_0^1 t^{h-1} (1-tre^{i\theta})^{-1/2} (1-tre^{-i\theta})^{-1/2} dt \qquad (25)$$

This last expression is a special case of the Appell function $F_1$. $F_1$ is one of the Appell hypergeometric function. In 1880, Paul Émile Appell has introduced a set of four hypergeometric functions $F_1, F_2, F_3, F_4$ that generalize Gauss's hypergeometric functions [29]. The function $F_1$ can be expressed by this integral [30]:

$$F_1(\alpha;\beta,\beta';\gamma;x,y) = \frac{\Gamma(\gamma)}{\Gamma(\gamma-\alpha)\Gamma(\alpha)} \int_0^1 t^{\alpha-1}(1-t)^{\gamma-\alpha-1}(1-tx)^{-\beta}(1-ty)^{-\beta'} dt$$

for $\text{Re}(\alpha)>0$ and $\text{Re}(\gamma-\alpha)>0$. The gamma function is represented by $\Gamma$.

With the properties of the $F_1$ Appell function, after some algebraic manipulations and by integration, we finally obtained this closed form of the function $v_2$:

$$v_2(r,\zeta) = r^{-h} \int_0^r \frac{\rho^{h-1}}{\sqrt{1-2\zeta\rho+\rho^2}} d\rho$$
$$= \frac{1}{h} F_1\left(h; -\frac{1}{2}, -\frac{1}{2}; h+1; r\zeta^\circ, \frac{r}{\zeta^\circ}\right)$$
$$+ \frac{2\zeta r}{h+1} F_1\left(h+1; \frac{1}{2}, \frac{1}{2}; h+2; r\zeta^\circ, \frac{r}{\zeta^\circ}\right)$$
$$- \frac{r^2}{h+2} F_1\left(h+2; \frac{1}{2}, \frac{1}{2}; h+3; r\zeta^\circ, \frac{r}{\zeta^\circ}\right)$$

(26)

where the variable $\zeta^\circ = \zeta + \sqrt{-1+\zeta^2}$ is introduced.

In addition, the analytical expression of the integral $\int_0^r \frac{\rho^{h-1}}{\sqrt{1-2\zeta\rho+\rho^2}} d\rho$ is given in terms of the Appell function. To our best knowledge, a closed form of this integral is described for the first time.

Now we calculate $v_1(r,\zeta)$:

$$\sum_{n=0}^\infty r^n P_n(\zeta) - h \sum_{n=0}^\infty r^n/(n+h) P_n(\zeta) = \sum_{n=0}^\infty n\, r^n/(n+h)\, P_n(\zeta)$$

Then

$$v_1(r,\zeta) = (1-2\zeta r+r^2)^{-1/2} - h\, v_2(r,\zeta) \qquad (27)$$

At the end and with the equation (20a),

$$v(r,\zeta) = (1-2\zeta r+r^2)^{-1/2} + (1/2 - h)\ v_2(r,\zeta) \tag{28}$$

Obviously if h=1/2 then the solution is the generating function.
Finally the convolution theorem (equation(18)) allows an integral representation of the solution with any prescribed function of the nonhomogeneous Robin conditions.

## 5. Some analytical solutions without the Appell functions

In addition to this general solution, the analytical solution could be written without the Appell hypergeometric functions if h is an integer :
h=j.

Gradshteyn et al [31] published the following indefinite integrals:
$\int P(r)\ (1-2\zeta r+r^2)^{-1/2}\ dr$ , where P(r) is a polynomial of some degree j.

For P(r)=1, the integral is:
$$\int (1-2\zeta\rho+\rho^2)^{-1/2}\ d\rho = Ln(2(r-\zeta+\sqrt{1-2\zeta\rho+\rho^2}\ ))$$

Let us write:
$$A_j = \int_0^r (1-2\zeta\rho+\rho^2)^{-1/2}\ \rho^j\ d\rho \tag{29}$$

$$A_0 = \int_0^r (1-2\zeta\rho+\rho^2)^{-1/2}\ d\rho = Log(\frac{r-\zeta+\sqrt{1-2\zeta\rho+\rho^2}}{1-\zeta}) \tag{30}$$

$A_1 = (1-2\zeta r+r^2)^{1/2} + \zeta\ A_0\ - 1$

$A_2 = 1/2\left((r+3\zeta)\ (1-2\zeta r+r^2)^{1/2} + (3\zeta^2-1)A_0\ -3\zeta\right)$

$A_3 = 1/3\left((r^2+5/2\ r\zeta+15/2\ \zeta^2-2)(1-2\zeta r+r^2)^{1/2} + (15/2\ \zeta^3 - 9/2\ \zeta)A_0\ - 15/2\ \zeta^2 + 2\right)$

with $m \geq 2$, $A_m = m^{-1}\left(R^{-1}\ r^{m-1} + (2m-1)\ \zeta \int_0^r \rho^{m-1}\ R\ d\rho - (m-1) \int_0^r \rho^{m-2}\ R\ d\rho\right) \tag{31}$

with $R=(1-2\zeta\rho+\rho^2)^{-1/2}$

With these equations and for h=j, it is very easy to express the solution without the Appell function.

## 6. The potential outside a sphere
$u_o(r,\zeta)$ is the potential outside a sphere with $r \geq 1$.
$\Delta u_o = 0$ (r>1).
The equation (6) becomes:

$\partial u_o(1, \zeta)/\partial r - h\, u_o(1, \zeta) = -f(\zeta)$

With the equation (4):

$$\underline{u}_{o\,(r,n)} = A_n\, r^n + B_n\, r^{-n-1} \qquad 0 \leq r \leq 1 \qquad (32)$$

Evidently, if $u_o$ is finite as $r \to \infty$ then $A_n=0$ and $B_n = \underline{f}_{(n)}/(n+h+1)$.
The variable $H=h+1$ is introduced and $h>0$ then $H>1$.

$$\underline{u}_{o\,(r,n)} = (1/r)\, \underline{f}_{(n)}\, r^{-n}/(n+H) = (1/r)\, \underline{u}_{(1/r,n)} \qquad (33)$$

The solution of the third problem for the exterior of a sphere is this simple result:

$$u_o(r, \zeta) = (1/r)\, u(1/r, \zeta) \qquad (34)$$

with "u" the solution for the potential inside a sphere (equation (29)) with $H=h+1$.

## 7. An introduction to the Legendre transform of the Helmholtz equation

With the azimuthal symmetry, the Legendre transform of the Helmholtz equation ($\Delta + k^2$; k is a constant $\neq 0$) is the following:

$$r^2\, \partial^2 \underline{u}/\partial r^2 + 2r\, \partial \underline{u}/\partial r + (k^2 r^2 - n(n+1))\, \underline{u} = 0 \qquad (35)$$

The solution of this spherical Bessel equation is [17, 21]:

$$\underline{u}_{(r,n)} = C_n\, J_{n+1/2}(k\,r)/\sqrt{(k\,r)} + D_n\, Y_{n+1/2}(k\,r)/\sqrt{(k\,r)}$$

where $C_n$ and $D_n$ are arbitrary constants. $J_q(z)$ and $Y_q(z)$ are respectively the Bessel function of the first kind and the second kind (the solution with the spherical Bessel function of the second kind being inadmissible as it would make $\underline{u}_{(r,n)} \to -\infty$ as $r \to 0$ then $D_n=0$).
Let us write the spherical Bessel function: $j_n(x) = (2/\pi)^{1/2}\, J_{n+1/2}(x)/\sqrt{x}$

$$\underline{u}_{(r,n)} = C_n\, (\pi/2)^{1/2}\, j_n(kr) \qquad (36)$$

By the same way of the section 3, the Legendre transforms for the Dirichlet conditions and for the Robin conditions are respectively:

$$\underline{u}_{(r,n)} = \underline{f}_{(n)}\, j_n(k\,r)/j_n(k) \qquad (37)$$

with $C_n = \underline{f}_{(n)}\, (2/\pi)^{1/2}/j_n(k)$

$$\underline{u}_{(r,n)} = \underline{f}_{(n)}\, (2/\pi)^{1/2}\, j_n(k\,r)/((k/2)\, j_{n-1}(k) - k\, j_{n+1}(k) + (h-1/2)\, j_n(k)) \qquad (38)$$

The solution for the diffusion-reaction equation ($\Delta - k^2$) with the Dirichlet condition is the following:

$$\underline{u}_{(r,n)} = \underline{f}_{(n)}\, \ddot{\imath}_n(k\,r)/\ddot{\imath}_n(k)$$

with $\ddot{\imath}_n(x)$ the modified spherical Bessel function of the first kind [28],
$\ddot{\imath}_n(r) = (2/\pi)^{1/2}\, I_{n+1/2}(r)/r^{1/2}$ where $I_m(r)$ is the modified Bessel function of the first kind.

Now the question is how to find the inverse Legendre transform of $j_n(k\,r)/j_n(k)$ or $\ddot{\imath}_n(k\,r)/\ddot{\imath}_n(k)$. With the generating function $R=\sqrt{(r^2+\rho^2 - 2r\rho \cos\theta)}$, the cylindrical Bessel functions $Z_v$ can be expressed as the following general infinite sum [16, 31]:

$$Z_w(\beta R)/R^w = 2^w\, \beta^{-w}\, \Gamma(w) \sum_{n=0}^{\infty} (n+w)\, J_{n+w}(\beta\rho)/(\beta\rho)^w\, Z_{n+w}(\beta r)/(\beta r)^w\, C_n^{(w)}(\cos\theta)$$

where $C_n^{(w)}$ is a Gegenbauer polynomial, $0<\rho<r$ and ß an arbitrary complex number.
A degenerate addition theorem [31] gives (with $i^2 = -1$):

$$e^{i\beta \rho \cos\theta} = (2\pi/(\beta\rho))^{1/2} \sum_{n=0}^{\infty} (n+1/2)\, i^n\, J_{n+1/2}(\beta\rho)\, P_n(\cos\theta)$$

and then

$$e^{i\beta \rho \cos\theta} = (2\pi)^{1/2} \sum_{n=0}^{\infty} (n+1/2)\, i^n\, j_n(\beta\rho)\, P_n(\cos\theta) \qquad (39)$$

$ï_n(\rho)$ is related to the spherical Bessel function of the first kind $j_n(i\rho)$ by [28]:
$j_n(i\rho)=i^n\, ï_n(\rho)$,  $j_n(-i\rho)=(-i)^n\, ï_n(\rho)$   and   $ï_n(i\rho)=j_n(\rho)$.

With $\beta=-i$ we get:
$$e^{\rho \cos\theta} = (2\pi)^{1/2} \sum_{n=0}^{\infty} (n+1/2)\, ï_n(\rho)\, P_n(\cos\theta) \qquad (41)$$

Finally, the inverse Legendre transform of $ï_n(kr)$ is $\left((2\pi)^{-1/2}\, e^{kr\cos\theta}\right)$.

These first steps show that the inverse transform of the solution (equation (37)) of the Helmoltz equation for the first boundary problem is far more difficult than the inverse transform applied to the Laplace equation.

## 8. Some inverse problems with the Robin conditions
### 8.1 Heat and mass transfer

For example, if we think of Newton cooling at r=1, we could consider a model of insufficient insulating condition in the thermal energy context. The constitutive law would be that the stationnary rate of heat loss flux density is proportional to the difference in temperature of the material ($u(1)$) and its surroundings ($u_s$). So, the flux is $J_{(r=1)} = g\,(u(1)-u_s)$ where g is a constant heat transfer coefficient. The value of g depends on the type of the both materials, the velocity of fluid flow, etc. Using Fourier's law, we would have

$du/dr_{(r=1)} + h\,(u(1) - u_s)=0$.

The Robin conditions appear in all transfer phenomena. "h" is equal to a dimensionless number in heat transfer, the Biot number (Bi). In mass diffusion processes it is the "mass transfer Biot number". These Biot numbers are very important in engineering particularly focusing on the dynamics at the interface between two different materials, such as the boundary of a solid particle submerged in a fluid. In the cases considered and in the steady state condition, the interface is stationary and there is no phase change or chemical reaction at the interface [22].

These Biot numbers are given by: $Bi = h°L/k$
where
$h°$ is the convective surface heat/mass transfer coefficient ($Wm^{-2}K^{-1}$).

L is a characteristic length. It is the typical length scale that heat in the solid particle must diffuse to get to the surface. For the sphere, the radius is the best scale and is already a dimensionless number (r=1).

k is the thermal/mass conductivity of the solid.

The Biot number should not be confused with the Nusselt number. If k is the conductivity of the fluid, then the dimensionless number is the Nusselt number. In the latter the transfer coefficient and the conductivity must relate to the same phase. The Biot number compares the relative transport resistances, internal and external: $Bi=(L/k)/(1/h°)=$ "internal diffusion resistance"/ "external convection resistance". Briefly, $Bi<<1$ means that the "external convection resistance" dominates the problem and that the well-known lumped system model could be used.

The surface heat transfer coefficient is a key parameter but is more difficult to measure than the conductivity [32]. Obsously one of the interest of analytical solution is the analysis of its derivative with respect to the particular parameter h in order to quantify the sensitivity and the stability of the process. Usually the convective heat/mass transfer is measured indirectly, as reported in the reviews for example in the domain of mathematical simulation studies for thermal food processes [33, 34]. When the surface is exposed to a moving fluid, heat transfer coefficients are difficult to obtain experimentally because their values depend strongly on many variables. In most practical situations, these convection problems are solved by using a single value of the surface heat transfer coefficient on the entire surface exposed to a moving fluid with homogeneous Robin condition. The solution of the Laplace equation (equations (26), (28) and (29)) with nonhomogeneous Robin condition corresponds to the steady state temperature (or concentration) of a single sphere with is exposed to linear transfer at its surface into medium whose temperature (or concentration) is proportional to f(cosθ). Our exact solution could help for the inverse estimation of this constant h° or the Biot number [32, 33]. The nonhomogeneous Robin condition appears when an additional heat flux [35] and/or a coefficient proportional to the unknown ambient temperature [36] is added on the boundary. We can also see the function f(cosθ) as a correction factor of the surface heat transfer coefficient (see its axisymmetric variation on figure 6 of [33]).

An extension of these inverse problems is the domain of simultaneous heat and mass transfer. Moisture and heat transfer occur in many processes related to hygroscopic materials, such as baking and drying of foods. During these processes, the heat transport into and out of a sphere's surface by convection and evaporation is what constitutes simultaneous heat and mass transfer. A better understanding of simultaneous heat and mass transfer, followed by the formulation of adequate mathematical models contribute to the optimization of these processes, and product quality improvement. Also generally accepted for foods is a surface boundary condition of the form given below [33]:

$-kA\partial T/\partial r = h°A (T-T_{fluid}) - q_{evaporation}$

with A area of the sphere.

where the term on the left side of this boundary condition refers to heat conducted from the outer surface to the inside of the body, the first term on the right side is heat penetrating from the surroundings to the solid body by means of convection, and the second term on the right side denotes heat of evaporation (defined as joule per second). Two inhomogeneous Robin conditions could be used for the mass transfer and for the heat transfer, then the two potentials (temperature and concentration) in the sphere are directly calculated.

**8.2 Optical tomography**

The third boundary problem with $\Delta u - k^2 u = 0$ arises in steady-state diffusion based optical tomography, where light propagation is modeled by a diffusion approximation where the absorption coefficient is very small compared with the diffusion coefficient [37]. u describes the photon density in the medium. The Dirichlet condition means that the medium around the body (here a sphere) is a perfect absorber, and then photons are absorbed when crossing the surface, so that outside the domain the photon density equals zero. A more realistic boundary condition is the homogeneous Robin condition [10].

The boundary condition could be approximated by [37]:

$u + 2 \kappa A \partial u/\partial r = 0$

with κ the diffusion coefficient and *A* a parameter governing the internal reflection at the boundary. We can use different approaches to derive *A* from Fresnel's law, and for n=1.4 we get *A*=3.25 or *A*=2.74 [37]. It is clear that the equivalent "optical Biot number" is not known.

There are two possibilities to model the light sources incident on the boundary: collimated sources or diffuse sources. The diffusion equation cannot describe correctly collimated sources by definition, so we can represent a collimated pencil beam by an isotropic source at a depth that is accurate at distances larger than the mean free path from this source [12]. Diffuse sources can be regarded as an inward directed diffuse photon current, distributed over the illuminated boundary segment. The inclusion of the source as a photon current through boundaries modifies the homogeneous Robin condition to a non homogeneous Robin condition :

$u + 2 \kappa A \partial u/\partial r = -f(\zeta)$

The head of some animals and of humans has a spherical geometry far from a semi-infinite geometry where we know solution [12]. Our result will be useful to analyse the effect of the boundary in optical tomography.

### 8.3 Detecting corrosion damage

Another class of inverse problems is to study some problems of identifiability arising in the domain of non-destructive evaluation [38, 39]. A sample is given that is marked by some imperfections due to various causes, which are located either on an inaccessible part of its boundary. If we consider the problem of detecting corrosion damage then the goal is to determine quantitative information about the corrosion that possibly occurs on an inaccessible part of the surface of a metallic specimen and on the 'accessible' part, electrostatic data are collected. The coefficient h, in the electrostatic context, represents the reciprocal of the surface impedance. It represents the corrosion damage, and classically, it is intepreted as a coefficient of energy exchange. Usually, it's a generalized Robin conditions where h is a function [38].

### 8.4 Altimetry–gravimetry boundary value problem

The main purposes of physical geodesy are the determination of the external gravity field and the geoid. The third geodetic boundary value problem has a special importance for physical geodesy as it constitutes the mathematical background in determining the ondulations of the geoid and the variation of the gravity field [40]; Traditionally, these tasks are handled by solving this third boundary value problem in which the input data are gravity anomalies on the surface of the Earth. An essential quantity that describes the Earth's gravity field is the gravity potential $W$. The gravity vector **g** is the gradient of $W$ (direction of the vertical or plumb line). The normal gravity field (potential $U$), a first approximation of the actual gravity field, is generated by an ellipsoid of revolution with its centre at the geocentre, called the reference ellipsoid (for example the WGS-84 ellipsoid). The surfaces U=constant are called normal level surfaces and the direction of the normal gravity vector is called the direction of the normal vertical. The difference between the gravity potential $W$ and the normal gravity potential $U$ is called the disturbing potential $T$.

Altimetry–gravimetry boundary value problem in spherical approximation has the form [40] :

$\nabla^2 T = 0$ outside the sphere

$L_{sea}(T) = -\partial T/\partial r = \delta g_{sea}$  and   $L_{land}(T) = -\partial T/\partial r - 2/R \; T = \Delta g_{land}$ on the sphere

where $\delta g_{sea}$ is the gravity disturbance at sea; $\Delta g_{land}$ is the gravity anomaly on land; R is the Earth's radius.

The spherical approximations will cause an error of the order of the Earth's flattening. To decrease the effect of the Earth's flattening, corrections due to the ellipticity of the boundary were applied to the Dirichled boundary value problem [41].

Obviously, there is no azimuthal symmetry but our analytical result for the potential outside a sphere (equation (43)) could give an averaged solution for this third geodetic boundary problem.

## 9. Numerical Applications

These inverse problems show that the dimensionless Robin number "h" is an important unknown boundary parameter in many domains. The equations (26-28) allow the study of the sensitivity to errors in boundary conditions. The derivative of the function $v$ with respect to "h" is the following sum of nine functions:

$$\begin{aligned}
\frac{dv}{dh} =\ & -\frac{1}{2h^2} F_1\left(h; -\frac{1}{2}, -\frac{1}{2}; h+1; r\zeta°, \frac{r}{\zeta°}\right) \\
& + \left(\frac{1-2h}{2h}\right) \left(D_a\left(F_1\left(h; -\frac{1}{2}, -\frac{1}{2}; h+1; r\zeta°, \frac{r}{\zeta°}\right)\right)\right. \\
& \left. + D_c\left(F_1\left(h; -\frac{1}{2}, -\frac{1}{2}; h+1; r\zeta°, \frac{r}{\zeta°}\right)\right)\right) \\
& - 2\zeta r \left(\frac{3}{2(1+h)^2}\right) F_1\left(h+1; \frac{1}{2}, \frac{1}{2}; h+2; r\zeta°, \frac{r}{\zeta°}\right) \\
& + 2\zeta r \left(\frac{1-2h}{2(1+h)}\right) \left(D_a\left(F_1\left(h+1; \frac{1}{2}, \frac{1}{2}; h+2; r\zeta°, \frac{r}{\zeta°}\right)\right)\right. \\
& \left. + D_c\left(F_1\left(h+1; \frac{1}{2}, \frac{1}{2}; h+2; r\zeta°, \frac{r}{\zeta°}\right)\right)\right) \\
& + r^2 \left(\frac{5}{2(2+h)^2}\right) F_1\left(h+2; \frac{1}{2}, \frac{1}{2}; h+3; r\zeta°, \frac{r}{\zeta°}\right) \\
& - r^2 \left(\frac{1-2h}{2(2+h)}\right) \left(D_a\left(F_1\left(h+2; \frac{1}{2}, \frac{1}{2}; h+3; r\zeta°, \frac{r}{\zeta°}\right)\right)\right. \\
& \left. + D_c\left(F_1\left(h+2; \frac{1}{2}, \frac{1}{2}; h+3; r\zeta°, \frac{r}{\zeta°}\right)\right)\right)
\end{aligned}$$

(42)

where $D_a()$ and $D_c()$ are respectively the derivatives with respect to $a$ and $c$ of $F_1(a;b,b';c;w,z)$.

The Figure 1 shows the evolution of the fundamental solution $v$ and its derivative (equation 42) with respect to the Robin parameter. The figures A1-E1 illustrate the changes in magnitude and in distribution of the fundamental solution. Evidently, the generating function contributes to the general form with the particular point {r=1, θ=0}. When h→+∞ the solution converges to the Dirichlet boundary conditions (the magnitude decreases e.g.), and when h→0, to the Neumann boundary conditions (the magnitude increases e.g.). The figures A2-E2 describe the modifications of the derivative of the fundamental solution with respect to the Robin parameter "h". For clarity, these figures only show the result for the half sphere (0≤ r ≤1).

The numerical data of Figure 1 (h ∈ {0.01, 0.05, 0.5, 5, 10}) are an example of the fundamental solution to recover the solute concentration or temperature or "photon concentration", etc… from boundary data and source measurement. In the case of small value of "h", the measurement uncertainty of this parameter affects a large volume of the sphere. When the Robin parameter is ≥ 10*(0.5), the derivative of the fundamental function is more and more flat (except in the domain near of the point {r=1, θ=0}).

When the Robin parameter is around 0.5, a rapid change of the solution (figures B1-D1) must be underlined. The analytical form of the function $v$ (equation 28) allows a direct analysis of these changes with the Robin parameter. Due to the sum of the generation function and the function $(1/2 - h) v_2$, $h=0.5$ could be considered as a tipping point. It is a significant result. Without closed-form expression, these changes could not be easily understood. These non-trivial results demonstrate the accuracy of the method and above all a better understanding of the behavior of the fundamental solution.

This numerical application of inverse theory for Robin boundary conditions quantifies and confirms the difficulty of the inverse determination of the Robin parameter or the Biot dimensionless number in heat and mass transfer. These numerical results exhibit the sensitivity to errors in the Robin boundary conditions, and the versatility of the method.

## 10. Conclusion

The Laplace equation for the axisymmetric sphere problem and the Robin conditions is solved by the method of integral transforms for the interior and the exterior of the sphere. This analytical solution is expressed with the Appell hypergeometric function $F_1$. Analytical method is to understand the physical effects through the model problem for example the Biot number. It is also useful to validate the numerical method. Analytical solutions will never go out of style because of the ongoing need for verification of numeric solvers and for use as direct solvers in support of experimental measurement. Moreover this solution could helped some inverse problems for example in heat and mass transfer or in optics.

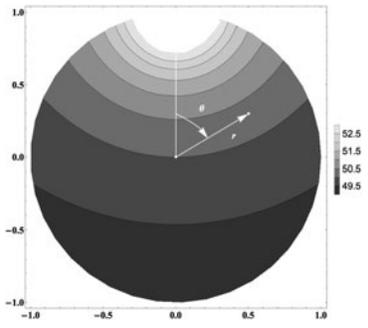 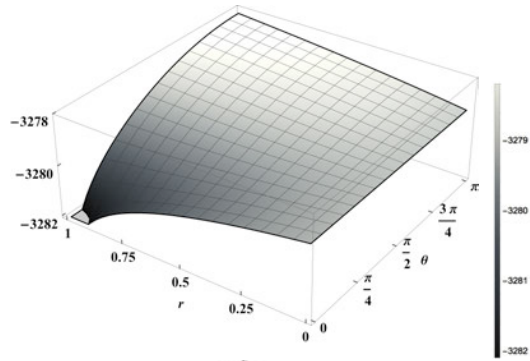

A1  A2

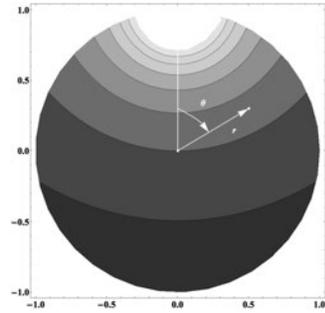 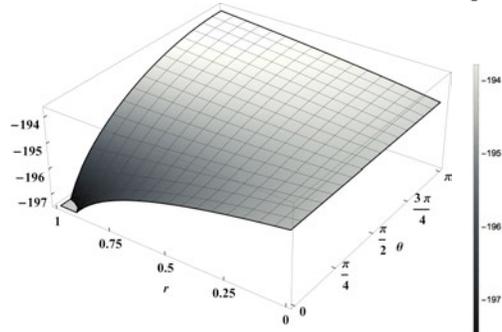

B1  B2

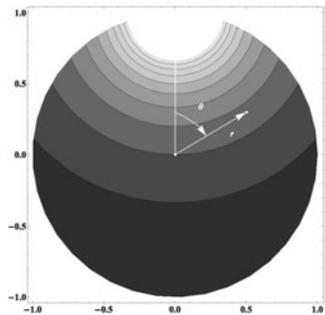 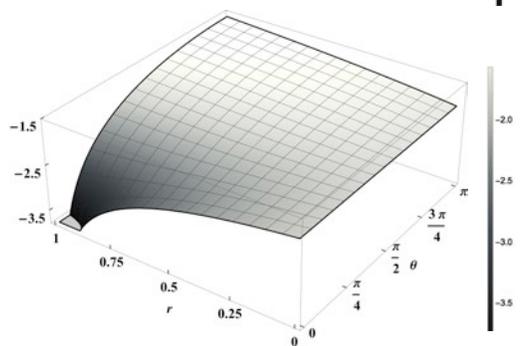

C1  C2

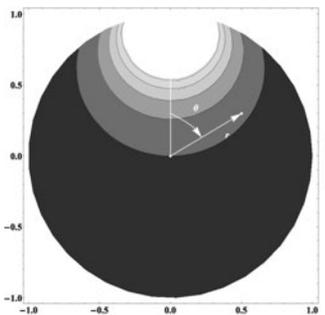 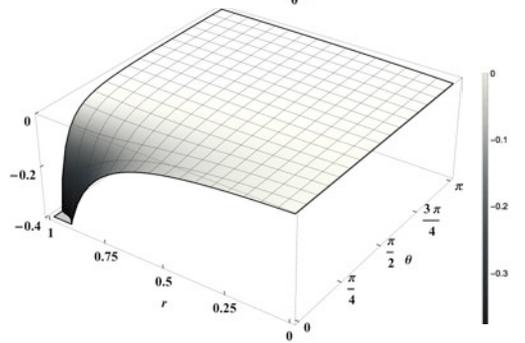

D1  D2

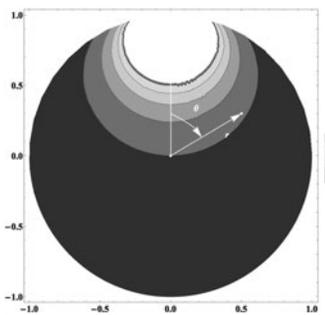 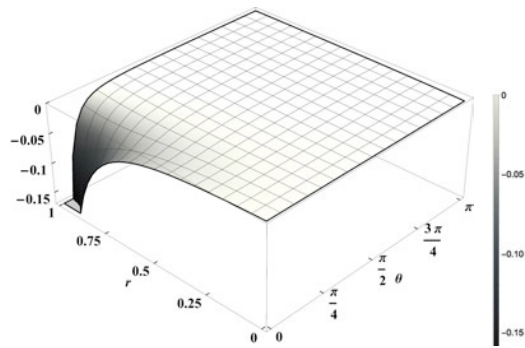

E1  E2

Figure 1. The evolution of the fundamental solution (equation 28) and its derivative with respect to the Robin parameter. The plots (A1-E1) correspond to the solution with h = 0.01, 0.05, 0.5, 5, 10. The plots (A2-E2) correspond to the derivative with h = 0.01, 0.05, 0.5, 5, 10.